# Superparamagnetic and metal-like Ru$_2$TiGe: a propitious thermoelectric material

Sanchayita Mondal[1,2], Krishanu Ghosh[3], R. Ranganathan[2], Eric Alleno[4], and Chandan Mazumdar[2]

[1] Maharaja Manindra Chandra College, 20 Ramkanto Bose Street, Kolkata 700003, West Bengal, India
[2] Condensed Matter Physics Division, Saha Institute of Nuclear Physics, 1/AF, Bidhannagar, Kolkata 700064, India
[3] Department of Physics, P.C. Vigyan College, Kathari Bag Road, Chapra, 841301, Bihar, India
[4] Univ Paris Est Creteil, CNRS, ICMPE, UMR 7182, 2-8, rue H. Dunant, F-94320 THIAIS, France
(Dated: September 18, 2022)

We report a study of structural, magnetic, heat capacity and thermoelectric properties of a Ru-based Heusler alloy, Ru$_2$TiGe. The magnetic measurements reveal that at higher temperatures, diamagnetic and Pauli paramagnetic contributions dominate the magnetic behaviour whereas, at lower temperatures (T ≤ 20 K), superparamagnetic interaction among clusters is observed. Effect of such magnetic defects is also evident in the electrical resistivity behaviour at lower temperatures. Though the temperature dependence of resistivity exhibits a metal-like nature, the large value of Seebeck coefficient leads to an appreciable power factor of the order of 1 mW/mK$^2$ at 300 K. Large power factor as well as low thermal conductivity results in a value of ZT = 0.025 at 390 K for Ru$_2$TiGe that is orders of magnitude higher than that of the other pure Heusler alloys and point towards its high potential for practical thermoelectric applications.

## I. INTRODUCTION

The power generation and refrigeration using thermoelectric materials are attracting increased global attention due to its environment-friendly technology as well as its long-term maintenance-free operation. The efficiency of a thermoelectric device to harvest waste heat depends on the figure of merit, ZT (ZT = $S^2T/\rho\kappa$, where S is the Seebeck coefficient, $\rho$ and $\kappa$ are the electrical resistivity and thermal conductivity, respectively), of the thermoelectric material used in the device. However, most of the materials, commercially available to use in the thermoelectric devices, are known to be rather toxic and expensive as well as not efficient enough, limiting their practical applications [1, 2]. Hence, to optimize the performance of a thermoelectric device, it is crucial to identify eco-friendly thermoelectric materials having large ZT. Consequently, many compounds, e.g., skutterudites [3], Clathrates [4] and Heusler alloys families [5, 6], etc. have been investigated to look for good thermoelectric materials. Amidst all different types of systems, the Heusler compounds have drawn great interests for the possibility to dope/substitute each of its constituents individually in order to optimize their physical properties [7].

Heusler alloys, X$_2$YZ (X /Y are transition metals, Z = p-block elements), crystallize in the cubic L2$_1$ structure having space group: Fm$\bar{3}$m. The L2$_1$ structure consists of four interpenetrating face-centered cubic (fcc) sublattices where the X -atoms occupy $(\frac{1}{4}\frac{1}{4}\frac{1}{4})$ and $(\frac{3}{4}\frac{3}{4}\frac{3}{4})$ positions, whereas Y and Z atoms are located at $(\frac{1}{2}\frac{1}{2}\frac{1}{2})$ and (0 0 0) positions, respectively [7, 8]. Interestingly, many different properties of Heusler alloys could be uniquely identified by their valence electron counts (VEC). Heusler compounds having VEC 24 have appeared to be promising for thermoelectric applications due to the presence of a narrow-or pseudo-gap in the vicinity of Fermi level (E$_F$). Starting with Fe$_2$VAl [9], many new Heusler alloys (Fe$_2$VGa, Fe$_2$TiSn etc. [10, 11]) having VEC 24 have been discovered and investigated till now in search of large ZT. For the same, one may have to keep in mind that a good thermoelectric material demands high thermopower and electrical conductivity as well as low thermal conductivity. The primary challenge one generally faces while identifying a good thermoelectric material is to decrease thermal conductivity and enhance electrical conductivity at the same time. It was found that in case of Fe$_2$VAl, the lattice thermal conductivity could be substantially reduced by doping/substitution with heavier element [12]. Extending this idea, one would therefore be tempted to use of heavier elements in identifying and synthesizing new Heusler materials, maintaining the valence electron count to be 24. As Ru is isoelectronic to Fe and have larger atomic size and weight, the structural formation of some Ru-based Heusler compounds having VEC 24 have been reported recently. Among those alloys, semi-metallic ground state has been observed in Ru$_2$NbAl [13], Ru$_2$NbGa [14] and Ru$_2$TaAl [15] whereas Ru$_2$VZ (Z = Al, Ga) [16–18] show metallic behaviour. Although having semi-metallic ground state, only Ru$_2$NbAl among the pristine compounds has shown an appreciable value of ZT at 300 K (ZT$_{300K}$ =0.0052) [13]. In spite of the fact that most of the Ru-based compounds exhibit relatively low thermal conductivity than that of the Fe-based Heusler alloys, the difficulties in achieving high ZT actually lie in their low values of the thermopower. Among these Ru-based alloys, a recent theoretical study has predicted Ru$_2$TiGe to be a potentially good thermoelectric material [19] as it showcases the possibility of having high thermopower as well as a semi-metallic ground state that are required for a good thermoelectric material. In comparison to Fe$_2$VAl, Ru$_2$TiGe contains two much heavier elements, viz., Ru and Ge, while the atomic weight of V and Ti are almost comparable.

In addition to their potential of exhibiting good thermoelectric behaviour, the Heusler alloys are also known to exhibit many intriguing magnetic properties. Interestingly, the total magnetic moment (M) per unit cell of a Heusler material can be estimated using Slater-Pauling rule: $M(\mu_B) = |VEC - 24|$ [20]. Thus, Heusler alloys with VEC 24 are expected to be nonmagnetic with a vanishing total magnetic moment per unit cell. Several materials having VEC 24 are indeed found to be nonmagnetic [9, 21, 22]. Nevertheless, close inspections in the experimental measurements of magnetic properties of these alloys reveal that many of the reported compounds could even be rather categorized as marginally magnetic and in some cases, the reported deviation in magnetic behaviour are also argued to have a number of different origins. For instance, while one study reported cluster glass behaviour for $Fe_2VAl$ [23], another study reveals the presence of superparamagnetism (SPM) in $Fe_2VAl$ [24]. SPM is also observed in other Fe-based and Ru-based Heusler alloys having VEC 24, viz., $Fe_2VGa$, $Ru_2NbAl$, etc. [10, 13]. Some other compounds in this class e.g., $Fe_2TiSn$, $Ru_2NbAl$ etc. [13, 25] show ferromagnetic interactions too. It is generally accepted that most of these various magnetic behaviours arise due to the structural antisite disorders that usually get developed during the synthesis and annealing process. Thus, a detailed study of the magnetic properties of $Ru_2TiGe$ could also help us in understanding the microstructural aspects that in turn can influence the thermoelectric properties.

In the present work, we accordingly report a detailed study on $Ru_2TiGe$ through structural, magnetic, heat capacity as well as thermoelectric measurements. It should be noted here that though any of the constituent elements are usually not found to exhibit magnetic ordering, our study reveals the presence of superparamagnetically interacting clusters in $Ru_2TiGe$. Thermoelectric measurements suggest that despite having metal-like ground state, a relatively large thermopower has been observed in this compound, leading towards a ZT value of ~0.025 at 390 K, the largest among the undoped Heusler compounds still reported.

## II. EXPERIMENTAL METHODS

A polycrystalline $Ru_2TiGe$ ingot was synthesized by arc melting technique under flowing Ar atmosphere. Stoichiometric amounts of the constituent elements Ru (>99.9%), Ti (>99.99%) and Ge (>99.9999%) were melted several times to achieve homogeneity. The weight loss in this process was observed to be less than 0.5%. The as-cast ingot was then wrapped in Ta-foil and annealed at 1273 K for 48 hours in a vacuum-sealed quartz tube followed by quenching in ice-water. The sample was annealed again at 1223 K for 12 hours after cleaning its surface and following the same quenching procedure. The ingots were then cut in appropriate shapes and polished and again annealed for 2 hours at 1173 K following the similar procedure in order to eliminate the surface strain that might have generated due to the mechanical stress in the process of cutting and polishing, as some Heusler alloys are observed to be highly sensible to such coldwork [26, 27]. The sample homogeneity and chemical composition of the annealed sample were evaluated by using the Wavelength Dispersive Spectroscopy based Electron Probe Micro-Analysis (WDS-EPMA) technique [Model: SX 100, M/s Cameca, France]. The powdered X-ray diffraction (XRD) technique was adapted to check the single phase nature of $Ru_2TiGe$ at 300 K using Cu $K_\alpha$ radiation in a powder diffractometer equipped with a rotating anode X-ray source at 9 kW [Model: TTREX III, M/s Rigaku Corp., Japan]. The XRD measurements were further carried out down to 12 K in order to look for any probable structural distortion. The Rietveld refinement technique by using FULLPROF software [28] is used to analyze the XRD patterns. Thermal transport [$\rho(T)$, $S(T)$, $\kappa(T)$], heat capacity [$C_P$], Hall coefficient [$R_H$] and magnetic [$M(T, H)$] properties were measured using commercial set ups [Models: PPMS Evercool-II and SQUID-VSM, M/s Quantum Design Inc., USA].

## III. RESULTS AND DISCUSSION

### A. Structural details

The XRD pattern of $Ru_2TiGe$ taken at 300 K is shown in Fig. 1. All the diffraction lines except a peak with negligible intensity (~ 1% of most intense peak) at $2\theta \sim 37°$, can be indexed considering the $L2_1$ crystal structure (Fig. 1). The lattice parameter estimated to be 6.034(1) Å, matches well with the earlier reported value [29]. The extra peak at $2\theta \sim 37°$, appeared due to the presence of a negligible amount of unreacted Ru, and have earlier been reported in many other Ru-based Heusler compounds [14, 29]. Another secondary phase was detected by EPMA (white areas in the inset of Fig. 1), where localized analysis found $Ru_{1.06}Ge_{0.94}$ as an apparent composition, most likely corresponding to the RuGe-Ru eutectic reported in the binary Ru-Ge phase diagram [30]. The absence of observed lines for RuGe in the XRD pattern most likely arises from its B2 crystal structure [31] and the value of its lattice parameter $a = 3.011$ Å close to half the lattice parameter of $Ru_2TiGe$ (6.034 / 2 = 3.017 Å, leading to a complete overlap of their respective lines. The presence of such small fraction of Ru as a secondary phase in these materials has been argued to have no influence in their transport and magnetic properties at all [14, 29]. Similarly, we argue here that a minute fraction of non-magnetic and metallic RuGe [31] does not affect the physical properties of $Ru_2TiGe$ either. It may also be noted here that the signature of structural antisite disorder is frequently observed in the Heusler compounds in perturbation to the ordered $L2_1$ phase. When Y and Z atoms exchange their respective positions and distribute evenly, the B2 type disordered state is resulted. The occurrence of B2 phase is generally reflected in the XRD pattern by diminished intensity of (111) diffraction line. In case of $Ru_2TiGe$ forming in the ordered $L2_1$ crystal structure, the intensity of (111) peak is calculated to be 2 % of the most intense line. In the present diffraction pattern, the measured intensity of the said diffraction line is ~1.9 % suggesting the sample is primarily forming in the ordered $L2_1$ phase. The elemental composition of the matrix, derived from the EPMA measurements is $Ru_{1.91(1)}Ti_{1.05(3)}Ge_{1.03(3)}$. This deviation from the stoichiometry is in agreement with the occurrence of Ru and RuGe secondary phases and also indicates the presence of $Ti_{Ru}$ and $Ge_{Ru}$ antisite defects where some of the Ti (Y) and Ge (Z) atoms occupies Ru (X) sites.

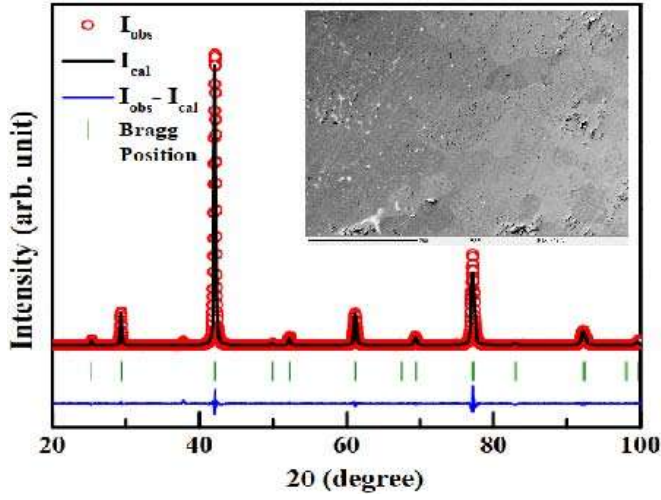
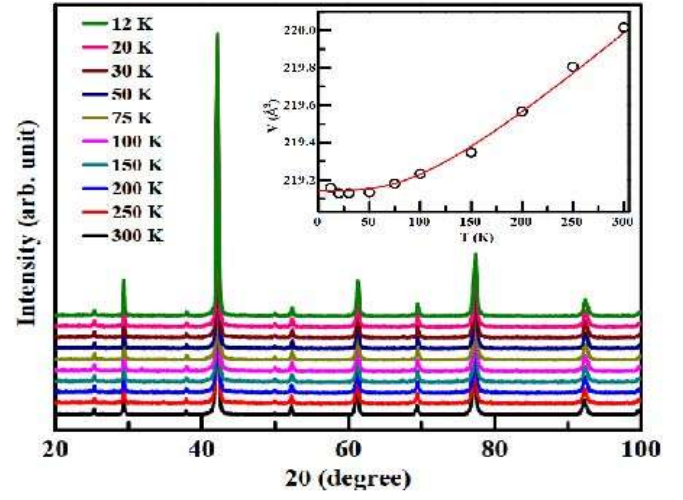

FIG. 1: Powdered X-ray diffraction pattern of $Ru_2TiGe$, measured at room temperature & indexed considering the $L2_1$ crystal structure; Inset: Back scattered image of $Ru_2TiGe$.

FIG. 2: Low temperature XRD pattern of $Ru_2TiGe$ down to 12 K; Inset: Temperature dependance of unit-cell volume of $Ru_2TiGe$. Solid line represents a fit to Eq. 1.

The XRD patterns collected at different temperatures between 300 K to 12 K do not show any detectable change in primary diffraction pattern-type indicating that the crystal structure remains unaltered down to 12 K, the lowest temperature achievable in our diffractometer (Fig. 2). The lattice parameter gradually decreases with decreasing temperature similar to that observed in most of the materials. The unit-cell volume of $Ru_2TiGe$ is plotted as a function of temperature and fitted (Fig. 2, inset) using the following equation [32]

$$V(T) = \gamma_G U(T)/K_0 + V_0, \quad (1)$$

where $V_0$ represents the unit-cell volume at T = 0 K, $K_0$ is the bulk modulus, and $\gamma_G$ is the Grüneisen parameter. U(T), the internal elastic energy can be described by considering the Debye approximation as

$$U(T) = 9NK_BT\left(\frac{T}{\Theta_D}\right)^3 \int_0^{\frac{\Theta_D}{T}} \frac{x^3}{e^x - 1} dx \quad (2)$$

where N is the number of atoms per unit cell and x = $\hbar\omega/k_BT$. From the fit of V(T) curve of $Ru_2TiGe$, the Debye temperature $\Theta_D$ and the Grüneisen parameter $\gamma_G$ are estimated to be 400 K and 1.6, respectively.

B. Magnetic properties

1. magnetic susceptibility

The temperature dependance of magnetic susceptibility [χ(T)] behaviour of $Ru_2TiGe$ has been investigated applying external magnetic field (H) of strengths 10 kOe and 70 kOe. The susceptibility values measured under zero field cooled (ZFC) and field cooled (FC) protocols at 10 kOe do not show any thermoremanence behaviour. The values of χ(T) at 10 kOe found to be very small though remain positive throughout the whole temperature region (Fig. 3, top). In contrast, the susceptibility values measured at 70 kOe become negative (Fig. 3, bottom) in the intermediate temperature range (100–200 K) suggesting the presence of diamagnetism in $Ru_2TiGe$. However, a minor upturn could be seen in the χ(T) curves for both the fields at higher temperatures, indicating Pauli paramagnetic (PPM) behaviour.

Experimentally observed χ(T) curves can be well explained considering a modified Curie-Weiss equation [33], expressed as)

$$\chi(T) = \frac{C}{T - \theta_p} + \chi_0 + \alpha T^2, \quad (3)$$

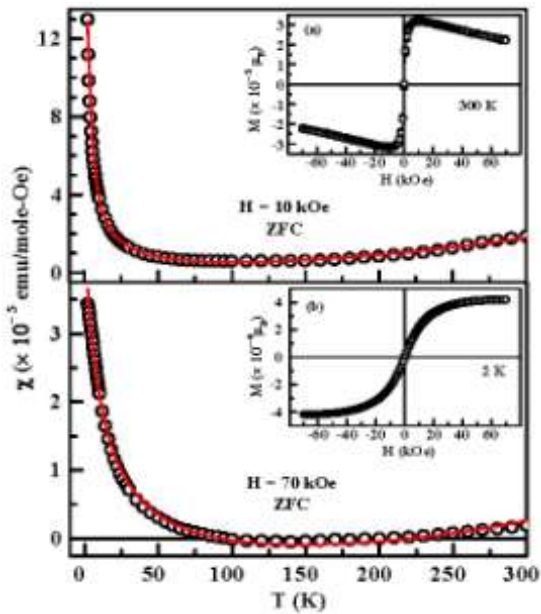

FIG. 3: Temperature dependence of magnetic susceptibility of $Ru_2TiGe$ measured at (top) H = 10 kOe and (bottom) 70 kOe under ZFC configuration along with a fit to Eq. 3; Insets: Isothermal magnetization at (a) 300 K and (b) 2 K of the same sample.

where the first term is the well known Curie-Weiss expression, $\chi_0$ is temperature independent diamagnetic and/or Pauli paramagnetic contributions and $\alpha T^2$ is the higher order term in the expanded Pauli paramagnetic equation which is generally neglected in the zeroth order approximation [33]. Using Eq. 3, the ZFC susceptibilities for both 10 kOe and 70 kOe applied field have been fitted and the fit parameters are shown in Table I. The values of $\chi_0$ for both the fields are found to be negative suggesting the predominant magnetic contribution is diamagnetic in $Ru_2TiGe$. However, the higher absolute value of $\chi_0$ at 70 kOe revels that the effect of Pauli paramagnet contribution has been suppressed on applying higher field and our estimated values of $\chi_0$ are the resultant of these two components ($\chi_0 = \chi_{PPM} - \chi_{dia}$). Apart from these two contributions, very small values of effective paramagnetic moment ($\mu_{eff}$) and paramagnetic Curie temperature ($\theta_P$) point towards the presence of another additional short range magnetic component, effect of which is evident in the relatively sharp upturn in $\chi$-T curves at low temperatures below ~ 25 K (Fig. 3). Such additional magnetic contribution may have been originated from the discernible deviation of stoichiometry that leads to small percentage of $Ti_{Ru}$ and $Ge_{Ru}$ antisite defects in this compound. It should be pointed out here that appearance of magnetic contribution from these defects is although rare, yet similar short range magnetization have earlier been reported in quite a few compounds e.g., $Ru_2NbAl$ [13], $SrRuO_3$ [34] etc., where none of the constituent elements are generally considered to be magnetic in nature. Effect of those defects and disorders in $Ru_2TiGe$ are further explored through isothermal magnetization measurements described below.

TABLE I: Parameters extracted from the fit of ZFC $\chi(T)$ curves for H = 10 kOe and 70 kOe to Eq. 3 for $Ru_2TiGe$.

| H (kOe) | $\chi_0$ (emu/mol-Oe) | $\mu_{eff}$ ($\mu_B$) | $\theta_p$ (K) | $\alpha$ (emu/mol-Oe-T$^2$) |
|---|---|---|---|---|
| 10 | $-0.75 \times 10^{-6}$ | 0.058 | $-1.32$ | $1.93 \times 10^{-10}$ |
| 70 | $-6.32 \times 10^{-6}$ | 0.069 | $-11.85$ | $0.78 \times 10^{-10}$ |

2. Isothermal magnetization

In order to further understand the magnetic behaviour of $Ru_2TiGe$, the isothermal magnetisation data have been collected at different temperatures as a function of externally applied magnetic field. The magnetic isothermal [M(H)] curve, taken at 300 K [Fig 3, inset(a)], increases initially upon increasing magnetic field up to 10 kOe after which the slope of the curve changes and becomes negative throughout the remaining magnetic field range scanned (10 < H ≤ 70 kOe). Such magnetic behaviour could be a result of simultaneous presence of two different type of magnetic contributions; one having positive magnetization and other with a negative component. Generally, negative magnetization has its origin in the diamagnetic contribution which varies linearly with increasing applied magnetic field, yielding a negative slope. Magnetic components which contribute towards the positive magnetization must possess a non-linear or a combination of both linear and non-linear variation with magnetic field, as only linear contributions can not results in the non-linear variation of the isothermal magnetization observed here. The positive contribution in this case increases rapidly with increasing applied field up to 10 kOe and then must have approached a saturating tendency suggesting that the positive contribution either could be of ferromagnetic type or a combination of paramagnetic and ferromagnetic interactions. In contrast, the isothermal magnetization at 2 K increases slowly with applied field and shows a approach-to-saturation-like behaviour [Fig 3, inset(b)]. In both the isotherms, taken at 300 K and 2 K, no hysteresis behaviour were observed. Since, signature of any ferromagnetic long range order could not be detected in the susceptibility behaviour of $Ru_2TiGe$, the presence of a subtle ferromagnetic contribution in the magnetic isotherm taken only at 300 K could possibly be generated from an impurity phase, present beyond the detection limit of both the XRD as well as EPMA analysis performed on this material. Elemental Fe could be a possible candidate for the impurity phase as it also saturates near 10 kOe though its saturation magnetisation, 2.2 $\mu_B$, is much higher than the observed value (~ $3\times10^{-5}$ $\mu_B$ at 10 kOe) at 300 K in Ru$_2$TiGe. Considering the whole positive contribution to the M(H) curve at 300 K is the manifestation of Fe impurity in Ru$_2$TiGe, the amount of Fe present in the system is estimated to be 0.0015 % or 15 ppm. To avoid such percentage of impurity, one need to prepare sample with all the starting elements having purity >99.9985%. Although for the synthesis of Ru$_2$TiGe, we have used Ti and Ge having purity >99.9985%, the purity of Ru was limited to 99.9%. Thus, 99.9% pure Ru available to us might easily be the source of the miniscule Fe impurity discussed above. Such a small impurity would also remain beyond the detection limit of measurements like XRD, EPMA, etc. employed in this study. However, it may be noted here that aforesaid impurity has a visible influence mostly at higher temperatures, where the positive paramagnetic contribution of Ru$_2$TiGe is quite low and the diamagnetic contribution dominates. In addition, this diamagnetic contribution also has a very weak dependence (negative dM/dH) of magnetic field. On the other hand, the magnetization of the FM impurity has a rather large positive dM/dH in the low field region, but the magnetization saturates as the magnetic field increases. As a result, the effect of FM impurity, though of very tiny fraction, has an overwhelming influence in the low field region. Once the FM contribution to isothermal magnetization saturates at higher field, the negative dM/dH of the diamagnetic contribution starts to manifest its domination. In contrast, at 2 K, the inherent magnetization value is quite large, increases as the magnetic field is raised and reaches toward a magnetization value ~ $4\times10^{-4}$ $\mu_B$ at 70 kOe, which is one order of magnitude higher than that of the FM impurity. Such larger magnetization values suppress the manifestation of magnetic impurity effect at the low temperature isothermal magnetization result. Similar presence of elemental magnetic impurity is also reported earlier in literature [35]. Such type of impurity, however, have very little effects in their transport properties. It may be pointed out here that the effect of such FM impurity does not have any significant influence on the $\chi$(T) data taken at 10 kOe and 70 kOe in the temperature range 2–300 K, presented in the previous section. The measured temperature range is well below the FM ordering temperature of Fe (1043 K) and both the applied magnetic fields (10 kOe and 70 kOe) falls within the magnetic saturation region of the isothermal magnetization characteristics of the Fe impurity. Thus the minute presence of the said Fe impurity can only contribute a very small positive temperature independent term (~$10^{-6}$ emu/mol-Oe) to the $\chi$(T) curves. The effect of removal of such constant term from each $\chi$(T) curves only results in a bit higher $\chi_0$ values than that given in Table I.

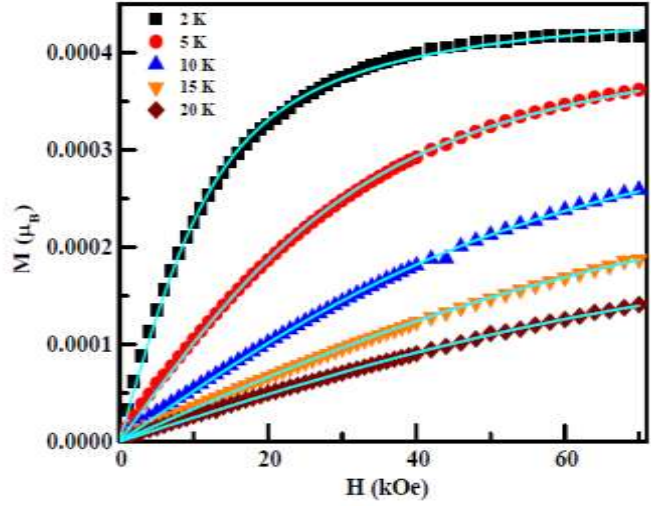

FIG. 4: Isothermal magnetization data at various temperature for Ru$_2$TiGe. The solid lines represent the fit of the data using Eq. 4.

To check the slope of the M(H) curve at intermediate temperatures, the data have also been taken at 20 K, 15 K, 10 K, and 5 K (Fig 4). It is observed that for all the temperatures the magnetic isotherms are nonlinear and show positive slopes. The nonlinearity increases as the temperature decreases. Since the susceptibility behaviour negate the possibility of the presence of any long range ordering in this compound, the deviation from the linear nature of the magnetic isotherms (expected in purely paramagnetic system) indicate towards the presence of a short range interaction. Generally, such S-shaped anhysteretic behaviour is observed in the materials having short range ferromagnetic interactions or a superparamagnetic (SPM) state.

To find the origin of the approach to such saturation-like behaviour, we have analyzed all the M(H) curves below 20 K considering the presence of SPM state in this material. To properly assign SPM to any system, two different conditions need to be simultaneously satisfied. Firstly, S-shaped anhysteretic curves should be described by a Langevin function and secondly, all the magnetic isotherms in the temperature range where the superparamagnetism is manifested, should overlap in a universal curve when plotted as a function of H/T [36]. We found that all the M(H)

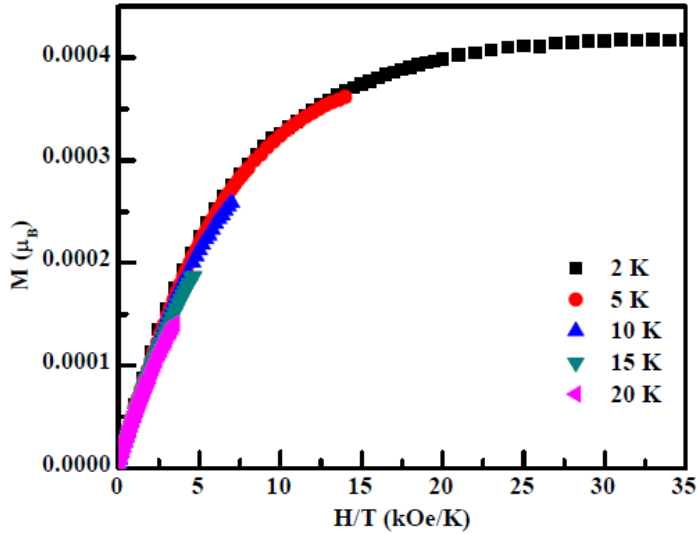

FIG. 5: The SPM state follows universal curve at low temperature isotherm for Ru$_2$TiGe.

curves below 20 K can be well-explained using the Langevin function [23, 24], given by

$$M(H) = M_S L(x) \qquad (4)$$

where x = M$_S$ is the saturation magnetization, μ represents the average magnetic moment per cluster and $\frac{\mu H}{k_B T}$, L(x)= coth(x) − 1/x, is the Langevin function. The parameters extracted from the fits are listed in Table II. It can be observed that for all the temperature in the region 20 K to 2 K, the magnetic moment per SPM cluster shows a little variation around ~6 μ$_B$. The saturation magnetization and the number of SPM cluster per mole are estimated to be ~ 4×10$^{-4}$ μ$_B$/f.u. and ~ 10$^{20}$/mole and are found to be close enough in the temperature range from 2–20 K. In order to check the other condition, all the magnetic isotherms below 20 K are plotted as a function of H/T (Fig. 5). Interestingly, it can be seen that they all indeed follow a single universal curve, confirming the presence of SPM in Ru$_2$TiGe below 20 K. It can be concluded from the analysis of the isothermal magnetization of Ru$_2$TiGe that the superparamagnetically interacting clusters develope below 20 K, whereas a competition between diamagnetic and ferromagnetic contributions is apparent at the higher temperatures.

TABLE II: Parameters extracted from the fit of magnetic isotherms of Ru$_2$TiGe to Eq. 4 for temperature range 2 - 20 K.

| T (K) | μ (μ$_B$) | M$_s$ (×10$^{-4}$μ$_B$/f.u.) | M$_s$/μ (×10$^{-5}$/f.u.) | N (×10$^{20}$/mole) |
|---|---|---|---|---|
| 2 | 5.29 | 4.61 | 8.71 | 0.52 |
| 5 | 5.16 | 4.54 | 8.80 | 0.52 |
| 10 | 5.94 | 3.96 | 6.67 | 0.40 |
| 15 | 6.80 | 3.33 | 4.90 | 0.29 |
| 20 | 2.14 | 2.58 | 2.14 | 0.12 |

C. Heat capacity

To ensure the veracity of the magnetization measurements, another independent experimental technique, the heat capacity [C(T)] has been carried out in the temperature range 2–300 K for Ru$_2$TiGe in absence of any external magnetic field (Fig. 6). The C(T) curve starts to increase as the temperature raises and does not show any signature of long range magnetic order supporting the results inferred from the magnetic measurements.

Generally, the observed heat capacity behaviour can be explained considering the Debye-Sommerfeld equation [37], given by,

$$C(T) = \gamma_S T + 9nR \left(\frac{T}{\Theta_D}\right)^3 \int_0^{\frac{\Theta_D}{T}} \frac{x^4 e^x}{(e^x - 1)^2} dx \qquad (5)$$

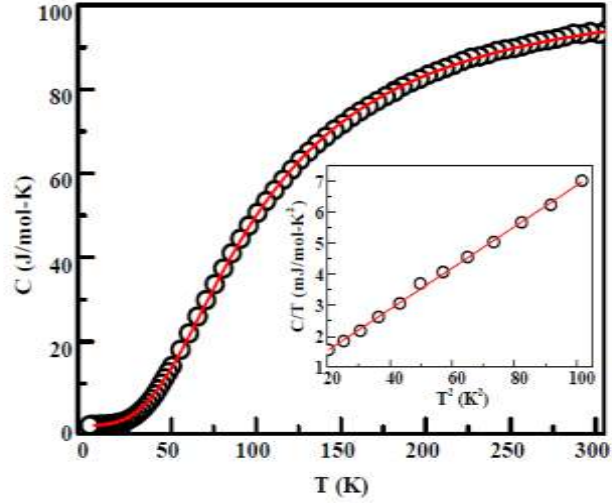

FIG. 6: Specific heat as a function of temperature of Ru$_2$TiGe. Solid line represents the fit to Eq. 5; Inset: C/T vs T$^2$ plot at low temperatures along with the fit to Eq. 6.

where $\gamma_S T$ is the electronic specific heat and the second term represents the lattice/phonon contribution to the heat capacity. The Sommerfeld coefficient $\gamma_S = \frac{1}{3}\pi^2 D(E_F) k_B^2$, where D(E$_F$) is the density of states at the Fermi level, E$_F$ and n represents the number of atoms per formula unit (for Ru$_2$TiGe: n = 4), Θ$_D$ is the Debye temperature and $x = \hbar\omega/k_B T$. This above discussed Debye-Sommerfeld model could reproduce well the heat capacity data in the temperature range 35–300 K. The value of Θ$_D$ and γ$_S$ estimated from the fit are 407 K and 7.6 mJ/mol-K$^2$ respectively. The value of Θ$_D$ is found similar to that obtained from the fit of lattice volume as a function of temperature (Sec. III A). A close inspection at the lower temperatures however reveals that the fit is not as good as it is in the higher temperatures and this is also reflected in the relatively high estimated value of γ$_S$ comparing standard metallic system [37]. This value of γ$_S$, obtained from the fitting of heat capacity data to Eq. 5, is not proper as γ$_S$ is generally a low temperature phenomenon [37] and thus a proper analysis of heat capacity data at lower temperature is crucial.

In the low temperature region Eq. 5 can be simplified as

$$C(T) = \gamma_S T + \beta T^3 + \delta T^5 + \cdots\cdots \quad (6)$$

where β, δ are the coefficients. Below ~ Θ$_D$/50, δT$^5$ and other higher order terms could be neglected and the heat capacity behavior in the representation of C/T vs. T$^2$ is expected to show a linear dependance. The Debye temperature can be estimated from the slope (β) of such straight line: $\beta = (12\pi^4/5)Nk_B/\Theta_D^3$. As expected in case of a standard material, the C/T vs. T$^2$ plot of Ru$_2$TiGe is found to be linear (Fig. 6, inset). The value of γ$_S$ and β are found to be 0.2 mJ/mol-K$^2$ and 0.067 mJ/mol-K$^4$ respectively. The value of Θ$_D$, calculated using obtained β is 487 K, found to be higher than that estimated from the whole temperature range fit. It is primarily because Θ$_D$s, extracted using Debye-Sommerfeld equation, are known to have a small temperature dependence [38, 39]. In the Debye model, it was assumed that only low-frequency modes of lattice vibrations i.e., acoustic modes contributes to the lattice specific heat for the large temperature region although this assumption is actually valid only at low temperatures. Therefore, at very low temperature, Θ$_D$ should be temperature independent. As the temperature increases, not only the acoustic modes, but optical modes of lattice vibrations also found to contribute to the lattice specific heat that in turn decreases the value of Θ$_D$ at higher temperature. Generally, the variations of Θ$_D$ remain within 10% around their mean value for most of the elements, though some exceptions, for instance, 20% variations have also been observed in case of Zinc, Cadmium etc. [38, 39]. For general use, the value of Θ$_D$ (Θ$_D$ = 407 K for Ru$_2$TiGe), estimated from fit of the C(T) data in the temperature range 2–300 K should have better acceptability as it is averaged over the entire temperature region. The low but non-zero value of γ$_S$ in Ru$_2$TiGe reveals that only a small number of free carriers are available for the electric conduction for this material, yielding a bad metal-like/semi-metallic ground state.

D. Transport properties

1. Resistivity

The electrical resistivity behaviour as a function of temperature [ρ(T)] of Ru$_2$TiGe has been studied in the temperature range 2–390 K, shown in Fig. 7. The ρ(T) curve exhibits a positive temperature coefficient of resistivity (TCR) over a large range of temperatures from 40 K to 390 K, suggesting metallic nature as earlier reported for this compound [40]. The magnitude of ρ at 390 K is 443 µΩ-cm whereas at 2 K it is found to be 362 µΩ-cm giving a residual resistivity ratio (RRR = $\rho_{390K}/\rho_{2K}$) value of ∼ 1.22. Despite having a metal-like resistivity, where resistance decreases with decreasing temperature, low value of RRR suggests Ru$_2$TiGe to be a bad metal. A close inspection at the lower temperatures reveals that the TCR changes its sign from positive to negative at around 40 K and remain negative down to lowest measured temperature. It may be noted here that in systems having high resistances, sharp upturn at low temperature may arise from different quantum interference effects such as weak-localization, electron-electron interactions etc. [41–43]. Though ρ(T) curve of Ru$_2$TiGe exhibits negative TCR below 40 K, the value of ρ changes at a very slow rate. Thus, the above mentioned processes may not be the primary basis of the negative TCR in Ru$_2$TiGe. We have discussed earlier that the structural analysis of Ru$_2$TiGe suggests small presence of Ti$_{Ru}$ and Ge$_{Ru}$ antisite defects. Such structural defects are often argued to be responsible for the localization of charge carriers [44]. Mott's variable range hopping (VRH) [45] of electrons between exponentially localized states is a mechanism that could explain both the negative TCR and high value of ρ in such systems at low temperature. In this mechanism, the conduction is expected to propagate via electrons, hopping between localized sites that are energetically closed but not necessarily close in space. The conduction behaviour of VRH in three dimensional systems can be expressed as

$$\rho(T) = \rho_0 exp\left[\left(\frac{T_0}{T}\right)^{1/4}\right] \quad (7)$$

where $\rho_0$ represents the residual resistivity and $T_0$, the activation temperature, depends on the localization length ($\xi$) as $\xi^{-3}$ [45]. The resistivity data of Ru$_2$TiGe has been plotted as ln(ρ) vs.T$^{-1/4}$ below 40 K (Fig. 7, Inset) and fitted using Eq. 7. In spite of having an excellent fit, the value of $T_0$ is found to be extremely low ∼7×10$^{-7}$ K than that of the values reported in the literature [13, 46] thus also ruling out the possibility of any significant role of VRH mechanism in this material.

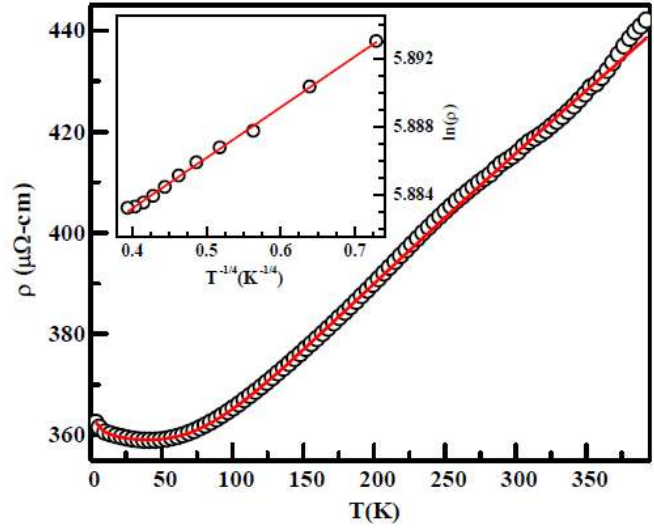

FIG. 7: Resistivity as a function of temperature for Ru$_2$TiGe with a fit to Eq. 9; Inset: Plot of ln($\rho$) vs. T$^{-1/4}$ below 40 K.

On the other hand, it may be noted here that the magnetic measurements discussed earlier suggest the development of superparamagnetically interacting clusters in the low temperature region in this material. In many non-magnetic systems containing a minute magnetic impurity, such resistivity minimum at low temperatures is found to be originated from the Kondo effect [47]. Recently, in some strongly correlated manganites as well as ferromagnetic metals, the Kondo effect is observed [48–50]. According to Kondo's theory, the resistivity below the minimum increases following the relation: $\rho_{Kondo} = \rho_0 - \rho_s \ln T$, where $\rho_0$ is the residual resistivity, the second term is the contribution from the interaction between the conduction electrons and the magnetic spins and $\rho_s$ represents the strength of the Kondo spin scattering [47]. Therefore, it appears that the negative TCR in Ru$_2$TiGe could be attributed to the formation of those magnetic clusters, pointing towards the Kondo-scattering mechanism. The ρ(T) behaviour above 40 K is almost linear and can be described using the Bloch-Grüneisen model [51] that considers the scattering of conduction electrons by the acoustic lattice vibrations. The temperature-dependent electrical resistivity of a metal according to the Bloch-Grüneisen model is given by

$$\rho(T) = \rho_{\Theta_D}\left(\frac{T}{\Theta_D}\right)^5 \int_0^{\frac{\Theta_D}{T}} \frac{x^5}{(e^x-1)(1-e^{-x})}dx \quad (8)$$

where, $\rho_{\Theta_D}$ is the value of resistivity at Debye temperature. In order to fit the ρ(T) curve for the entire temperature

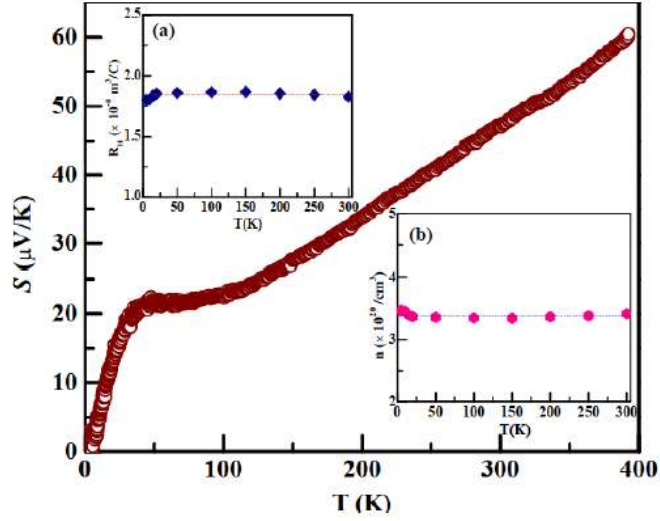

FIG. 8: Temperature dependance of Seebeck coefficient of Ru$_2$TiGe. Insets: (a) Hall coefficient and (b) hole concentration as a function of temperature

region, the following equation has been used combining both Kondo effect and the Bloch-Grüneisen formula

$$\rho(T) = \rho_0 - \rho_s \ln T + \rho_{\Theta_D} \left(\frac{T}{\Theta_D}\right)^5 \int_0^{\frac{\Theta_D}{T}} \frac{x^5}{(e^x - 1)(1 - e^{-x})} dx \qquad (9)$$

The ρ(T) curve for Ru$_2$TiGe fitted very well using the above equation (Fig. 7). The parameters extracted from the fit are $\Theta_D$ = 508 K, $\rho_{\Theta_D}$ = 110 μΩ-cm, $\rho_0$ = 361 μΩ-cm and $\rho_s$ = 1.1 μΩ-cm. It is important to note here that the values of $\Theta_D$, estimated from the resistivity measurement, found to be higher than that of the values extracted from the lattice thermal expansion and heat capacity data. We have already discussed that both longitudinal as well as transverse acoustic lattice vibrations were considered in the Debye model to calculate lattice specific heat theoretically. In contrast, only longitudinal phonons have been considered in the theory to develop Bloch-Grüneisen formula [38]. As the basic assumptions behind the Bloch-Grüneisen model and the Debye theory are quite different, consequently, the Debye temperature evaluated from electrical resistivity measurement differs from the $\Theta_D$, obtained by the heat capacity data. However, the $\Theta_D$ (400 K), evaluated from lattice thermal expansion is found to be close enough to that of the heat capacity. It is important to note that the thermal expansion parameter ($\Theta_D$) exhibit a Debye behaviour in the entire temperature region as the main contributions is provided by the phonons to thermal expansion. The coefficient of thermal expansion, α, is related to the heat capacity C and bulk modulus $K_0$ by α = $(1/V)(\partial V /\partial T)_P$ = $\gamma C/K_0 V$, where γ is Grüneisen parameter [32]. Generally, γ and $K_0$ have very weak dependence on temperature, the plot of α as a function of temperature essentially follows a Debye behavior. Because of these, the value of $\Theta_D$ estimated from the analysis of different data often found to differ in different analysis [38, 39]. Among all three $\Theta_D$ values of Ru$_2$TiGe, $\Theta_D$ = 407 K estimated from the heat capacity data is considered as the best representative of $\Theta_D$ in this system since the approximations considered in the Debye model, used to explain the temperature variation of heat capacity are most inclusive in nature.

2. Seebeck coefficient

In order to explore the thermoelectric properties of Ru$_2$TiGe, the temperature dependence of Seebeck coefficient [S(T)] has been investigated in the temperature range 2–390 K, as shown in Fig. 8. The value of Seebeck coefficient at 300 K is 47 μV/K, found to be higher than that of other Heusler alloys having VEC 24 like Fe$_2$VAl (∼35 μV/K), Fe$_2$VGa (∼30 μV/K), Ru$_2$NbAl (∼22 μV/K) and Ru$_2$NbGa (∼20 μV/K) [13, 46, 52–56]. The magnitude of the S(T) remains positive throughout the entire temperature region examined suggesting that the majority carriers for the thermoelectric transport must be holes in Ru$_2$TiGe. As the temperature raises, S(T) increases gradually and after exhibiting a hump-like shape with a maximum at T ≈ 50 K, it varies almost linearly above 120 K and attains a value of 60 V/K at 390 K. The S(T) onset of linear variations does not coincide with the onset of monotonous increase of the resistivity at around 80 K (Fig. 7). The maximum in S(T) rather matches with the lattice thermal conductivity maximum at 50 K (Fig. 9) discussed later. The appearance of the hump-like shape in S(T) curve of Ru$_2$TiGe can therefore be ascribed to the phonon drag effect, similar to that reported earlier in Fe$_2$TiSn [57]. However, the linear nature of S(T) curve at higher temperatures is in sync with observed metallic electric conduction behaviour of Ru$_2$TiGe.

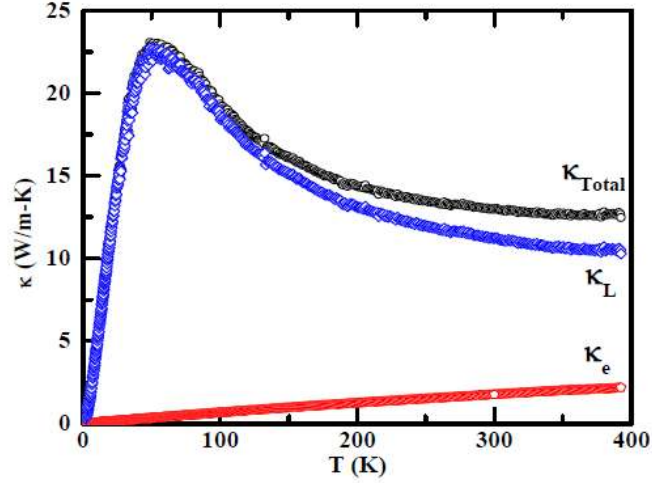

FIG. 9: Temperature variations of the total thermal conductivity $\kappa$, lattice thermal conductivity $\kappa_L$, and electronic thermal conductivity $\kappa_e$ for $Ru_2TiGe$ as a function of temperature.

3. Hall coefficient

The Hall coefficients ($R_H$) were measured for $Ru_2TiGe$ in the temperature range 5–300 K for externally applied magnetic field of 50 kOe in order to check the sign of the Seebeck coefficient already discussed. The Hall resistivity increases with rising magnetic field almost with a constant slope for all the investigating temperatures and the value of $R_H$ has been calculated using that constant slope. The magnitude of $R_H$ remains positive and almost temperature independent for the entire temperature regime (Fig. 8, Inset:(a)) indicating the dominance of hole-type carrier for thermoelectric transport in conformity with the positive sign of S(T) observed in this compound (Fig. 8). The value of observed $R_H$ is found to be $\sim 10^{-8}$ $m^3$/C, which is 100–1000 times larger than that of a conventional metals and nearly close to the elemental semimetals such as Sb. Assuming the presence of only one type of carrier i.e. holes, the hole concentrations(n) are calculated $\sim 3.3 \times 10^{20}$ $cm^{-3}$ (Fig. 8, Inset:(b)) using the relation n =1/$eR_H$, where e is the charge of electron. This estimated n value is subsequently 100–1000 times lower than a conventional metal and such small numbers of carriers are responsible for its low RRR and large thermopower.

4. Thermal conductivity

To further evaluate the thermoelectric performance of $Ru_2TiGe$, thermal conductivity ($\kappa$) was measured in the temperature range 2–390 K (Fig. 9). At low temperatures, $\kappa$ starts to increase rapidly with temperature and shows a sharp peak at ~50 K. Above 50 K, the value of $\kappa$ decreases and attains 12.4 W/m-K at 390 K (Fig. 9). For ordinary metals and semimetals, the total thermal conductivity is a sum of lattice ($\kappa_L$) and electronic ($\kappa_e$) contributions. The electronic thermal conductivity can be extracted using the Wiedemann-Franz law $\kappa_e\rho/T = L_0$, where $\rho$ represents the measured dc electric resistivity and $L_0 = 2.45\times10^{-8}$ $W\Omega K^{-2}$ is the Lorenz number. The lattice thermal conductivity $\kappa_L$ can then be evaluated by subtracting $\kappa_e$ from the observed $\kappa$. The estimated $\kappa_e$ is found to be only a small contribution to $\kappa$ (Fig. 9), indicating that the thermal conductivity of this material is essentially originated from $\kappa_L$ (Fig. 9). At low temperatures, $\kappa_L$ increases with temperature and a sharp maximum appears around 50 K because of the reduction in thermal scattering at low temperatures. The height of this peak in $\kappa_L$ commonly represents the degree of crystallographic order in a compound and for a structurally disordered system this peak in $\kappa_L$ remains absent [14, 58]. Thus, the sharp peak in $\kappa_L$ for the present compound agrees with its ordered $L2_1$ structure. The $\kappa_L$ value of $Ru_2TiGe$ at 300 K is estimated to be 11.2 W/m-K (Fig. 9), which is however lower than those in $Fe_2VAl$ (~ 28 W/m-K) and $Fe_2VGa$ (~ 17 W/m-K) [53–56]. This reduction may be attributed to smaller Debye temperature and speed of sound caused by substituting Fe by heavier Ru atom. Despite having a metal-like ground state, appreciably large value of thermopower as well as low $\kappa$ in $Ru_2TiGe$ increase the possibility of large ZT in this compound.

In order to understand the influence of various phonon scattering mechanism on the lattice thermal conductivity of $Ru_2TiGe$, the $\kappa_L(T)$ data was analyzed using the Debye-Callaway model [59, 60]. According to this model, the lattice thermal conductivity as a function of temperature is described by the following equation:

$$\kappa_L(T) = \frac{k_B}{2\pi^2 v}\left(\frac{k_B T}{\hbar}\right)^3 \int_0^{\frac{\theta_D}{T}} \frac{x^4 e^x}{\tau_P^{-1}(e^x - 1)^2} dx \qquad (10)$$

where x = , $\omega$ and $\theta_D$ are the phonon frequency and the Debye temperature, respectively, and v represents the average phonon velocity that is approximately equal to the speed of sound in the studied material. The phonon scattering relaxation time, $\tau_P$, is considered as the sum of different phonon scattering mechanism, defined as

$$\tau_P^{-1} = \frac{v}{L} + A\omega^4 + B\omega^2 T e^{-\theta_D/3T} + C\omega + D\omega^2 \qquad (11)$$

where L is the characteristic length defining the sample size and A, B, C and D are the free fitting parameters. The terms in the right of Eq. 11 are the scattering rates of phonon by the sample boundaries (SB), the point or mass defects (PD), the phonons (PP), the grain boundaries (GB) and the electrons (EP), respectively [60–62]. The frequency dependent expression ($C\omega$) for the phonon scattering rate by the grain boundaries has been included in Eq. 11 since $\kappa_L$ shows a $T^2$ variation (Fig. 10) at very low temperature in $Ru_2TiGe$. Such $T^2$ variation of $\kappa_L$ for silicon below ~70 K had previously been explained while deriving the above mentioned frequency dependent term $C\omega$ [61]. A metal-like ground state and the hole concentration of ~ $3.3 \times 10^{20}$ cm$^{-3}$) in $Ru_2TiGe$ compel to adapt the term $D\omega^2$ considering the phonon scattering rate by electron [62]. To fit the $\kappa_L(T)$, the value of $\theta_D$ = 407 K estimated from the specific heat data is utilized and v = 3200 ms$^{-1}$ is taken as the average speed of sound. Here, L = 2 mm was held as a fixed parameter giving its weak influence on the results.

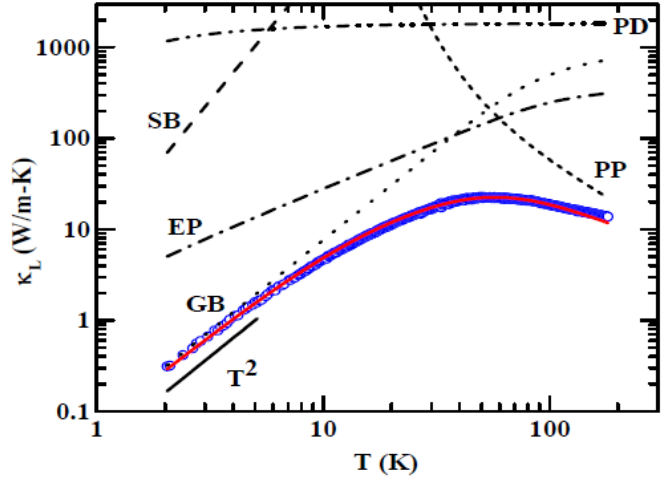

FIG. 10: Temperature variations of lattice thermal conductivity $\kappa_L$ for $Ru_2TiGe$. The open circles are the experimental data; the solid line is the fit to Eq. 10; the dashed curves are the theoretical limits on the lattice thermal conductivity imposed by different phonon scattering processes.

A good agreement between the fitted curve and the measured values can be seen in Fig. 10. The deviation at the high temperature region most likely arises from radiation artifacts on the measurement. The fitted parameters are presented in Table III. The relative weight of each scattering process on the lattice thermal conductivity can be better understood by replacing $\tau_P^{-1}$ in Eq. 10 by the individual scattering rate, yielding a theoretical limit imposed by each process on the lattice thermal conductivity. It can be interpreted from Fig. 10 that scattering by the sample boundaries and by the mass or point defects play a minor role in the $\kappa_L$ of $Ru_2TiGe$. At low temperature, grain boundary scattering dominates while electron and phonon scattering play a major role at intermediate and high temperature regions, respectively.

TABLE III: Parameters of the Debye-Callaway model fitted to the lattice thermal conductivity data in $Ru_2TiGe$. See main text for their definition.

| A($s^3$) | B(s/K) | C | D(s) |
|---|---|---|---|
| $5.7 \times 10^{-43}$ | $5.5 \times 10^{-18}$ | $3.6 \times 10^{-4}$ | $3.3 \times 10^{-17}$ |

### 5. Figure of merit

Using all three measured thermoelectric parameters i.e., Seebeck coefficient [S(T)], electrical resistivity [ρ(T)] and thermal conductivity [κ(T)], the value of the power factor (PF=$S^2$/ρ) and the dimensionless figure of merit (ZT =$S^2$T/ρκ) are estimated and illustrated in Fig. 11. Although $Ru_2TiGe$ exhibits metal-like electric conduction, its comparably large Seebeck coefficient leads to achieve higher value of power factor (1 mW/mK$^2$ at 300 K) for this material than that of the existing Fe-based and other Ru-based stoichiometric Heusler alloys. For example, the room temperature PF values of 0.6 and 0.1 mW/mK$^2$ have been reported earlier in literature for $Ru_2NbAl$ [13] and $Fe_2VAl$ [58], respectively. With the rise of temperature, PF also increases and reaches at a value of 2 mW/mK$^2$ at 390 K. Such PF value for present compound is of the same order of magnitude to $Fe_2VAl$-based and $Fe_2VGa$-based Heusler alloys that are suitably doped for the improvisation of their ZT values [48, 63, 64]. Similar to power factor, the estimated value of ZT at 300 K ($ZT_{300K}$ = 0.012) is also higher than that of $Ru_2NbAl$ for which largest value of ZT (0.0052 at 300 K) was obtained among all reported pristine Heusler compounds [13]. The ZT for $Ru_2TiGe$ is orders of magnitude higher than that of the most investigated non-doped thermoelectric material $Fe_2VAl$. Temperature dependance of ZT also exhibit a similar trend as observed for the power factor and yields almost two times higher value of 0.025 at 390 K than at 300 K. Nonetheless, it is still one order of magnitude smaller than that of the state-ofthe-art thermoelectric material at 300 K, $Bi_2Te_3$ which displays ZT = 0.8 at 300 K. Enhancement in ZT values are earlier reported by suitable doping/substitution in proper site of the pristine Heusler alloys viz. $Fe_2VAl$, $Ru_2NbGa$, $Ru_2TaAl$ etc. [15, 64, 65]. It may be noted here that a 20-times increased ZT value (0.0036) at 300 K was reported in $Fe_2VAl_{0.9}Si_{0.1}$ [58] than in non-doped $Fe_2VAl$ and by replacing heavier Ge atom in place of Si in $Fe_2VAl_{0.9}Si_{0.1}$, a total 722-fold increased ZT value of 0.13 at 300 K was achieved in $Fe_2VAl_{0.9}Ge_{0.1}$ [54]. Interestingly, being only a pristine compound, $Ru_2TiGe$ exhibits such a large value of ZT at room temperature, further improvement in ZT for $Ru_2TiGe$ is highly anticipated by following similar strategy through doping to alter the metal-like ground state to a semiconductor-like one by adjusting the charge carrier concentration and substitution to reduce the lattice thermal conductivity.

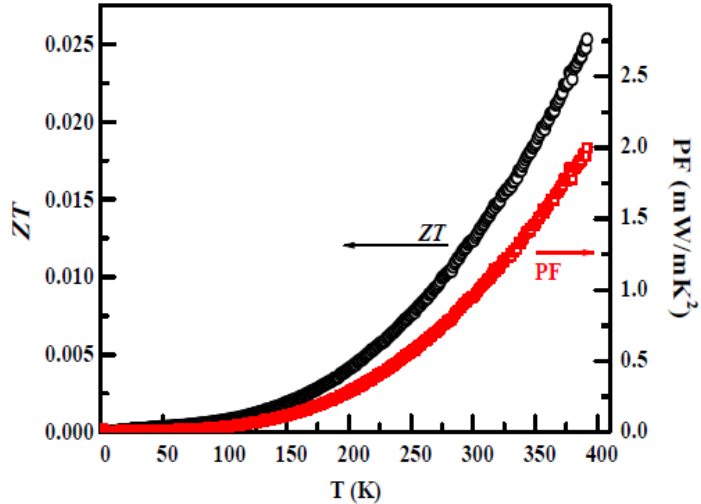

FIG. 11: $ZT$ and power factor as a function of temperature for $Ru_2TiGe$.

IV. CONCLUSION

An elaborate study on a Ru-based Heusler alloy, $Ru_2TiGe$, is performed through structural, magnetic, heat capacity and thermoelectric properties measurements. The magnetic properties at higher temperatures are dominated by temperature independent diamagnetic and Pauli paramagnetic contributions. In contrast to the observed magnetic behaviour at higher temperatures, superparamagnetic interaction among clusters is observed at lower temperatures below 20 K, despite the fact that any of the constituent atoms in $Ru_2TiGe$ is usually not known to order magnetically in nature. This short range magnetic interaction may have its origin in the small off-stoichiometry, noticed in otherwise structurally ordered (in $L2_1$ phase) $Ru_2TiGe$. Effect of such magnetic defects is manifested in the electrical resistivity behaviour at lower temperatures. Larger Hall coefficient ($R_H \sim 10^{-8}$ m$^3$/C), smaller Sommerfeld coefficient ($\gamma_S$ = 0.2 mJ/mol-K$^2$) as well as lower RRR ($\rho_{390K}/\rho_{2K}$ =1.22) than a conventional metal categorize $Ru_2TiGe$ as a bad metal or a semi-metal. In spite of having a metal-like resistivity, a large value of Seebeck coefficient in this compound results in a larger power factor of the order of 1 mW/mK$^2$ at 300 K than that of the other reported non-doped Heusler alloys. Large power factor as well as low thermal conductivity which is the result of employing heavy constituent elements like Ru and Ge in $Ru_2TiGe$, give rise to a superior value of ZT = 0.012 at 300 K in comparison to those reported in other pure Heusler alloys and point towards its high potential for practical thermoelectric applications. We have already discussed in the previous section that a many-fold increment in ZT can be achieved with proper doping and/or substitution and following similar strategy, great improvement in the thermoelectric performances of $Ru_2TiGe$ is also highly anticipated.